\documentclass[%
 reprint,
amsmath,amssymb,longbibliography,aps,prx,]{revtex4-2}
\pdfoutput=1
\usepackage{graphicx,multirow}%
\usepackage{amsmath}
\usepackage{dcolumn}
\usepackage{bm}
\usepackage{xcolor,colortbl}
\usepackage{color}
\definecolor{blue}{rgb}{  0,  0,    1}
\definecolor{d4blue}{rgb}{  0,  0.4470,    0.7410}
\definecolor{d12orange}{rgb}{0.8500    0.3250    0.0980}
\definecolor{g8yellow}{rgb}{0.    0.6    0.298}
\definecolor{ppurple}{rgb}{0.4940    0.1840    0.5560}

\newcolumntype{k}{>{\columncolor{blue!20}}c}
\newcolumntype{r}{>{\columncolor{blue!10}}c}
\newcolumntype{d}{>{\columncolor{red!10}}c}
\usepackage{comment}

\usepackage{hyperref}

\begin{document}

\preprint{APS/123-QED}

\title{``Universal" $\Delta_{max}/T_c$ in Fe-based Superconductors}

\author{ Siddhant Panda and  P. J. Hirschfeld}
\affiliation{
Department of Physics, University of Florida, Gainesville, Florida 32611, USA
}%

\date{\today}

\begin{abstract}
Iron-based superconductors display a large degree of variability in electronic structure at the Fermi surface, resulting in superconducting gap structures, $T_c$'s, and other properties that vary considerably from family to family.  Recently it was noted that across many different families of Fe-based systems, the ratio $\Delta_{max}/T_c$ is found to be quasi-universal with a large value of $\sim$3.5 compared to the the one-band BCS weak-coupling result of 1.76. Here $\Delta_{max}$ is the measured maximum gap across the Fermi surface. This remarkable fact was attributed to strong-coupling effects arising from Hund's metal physics.  Here we perform  a ``high-throughput" scan across band masses, { Fermi energies}, and interaction parameters in a weak-coupling Suhl-Matthias-Walker model.  We find that unexpectedly large values of $\Delta_{max}/T_c$ can be achieved within weak coupling, and that  quasiuniversal behavior similar to experiment emerges for those systems where interband interactions dominate intraband ones. However, within the current framework, a large mass contrast between bands is required.  
\end{abstract}

\maketitle


\section{\label{sec:intro} Introduction}

Iron-based superconductors have fascinated the superconductivity community for more than a decade.  Originally hailed after their discovery in 2008\cite{Kamihara08} as a second class of high-temperature superconductors similar to the cuprates, it was quickly realized that they display a much richer variability of superconducting behavior, due to their multiband, multiorbital character\cite{Hirschfeld_review}.  It is  widely believed these Fe-based superconductors
(FeSC) are unconventional, i.e. that pairing is  driven by repulsive interactions \cite{Fernandes22}.  Unlike cuprates, where a $d$ wave pair state is realized, most iron-based materials are thought to pair in a dominant $s$-wave state,  with different order parameter signs  on different bands \cite{Mazin08,Kuroki09}.  In addition, experiments show that significant gap anisotropy exists in some materials, including those with gap nodes.  Within the usual spin fluctuation theory, this gap structure is driven by several features of the effective interaction, but in particular by the distribution of orbital weights on each Fermi surface sheet\cite{Maier09,JZhang09,FernandezMartin2021}.

In addition to strongly differing low-temperature gap sizes and structures, the  the FeSC display a wide range of superconducting transition temperatures, from a few Kelvin up to $\sim 70$K in monolayer FeSe grown on SrTiO$_{3}$.   Variations in overall strength of the superconductivity across this family are generally attributed to Fermi surface features\cite{Graser09,Maier09,JZhang09,Ikeda2010,HKM2011,FernandezMartin2021} such as the number, size, and $d$- orbital content of the small Fermi surface pockets around the high symmetry points of the Brillouin zone, degree of nesting of the pockets, and Fermi velocities. Bare effective Coulomb interactions also vary substantially from material to material\cite{Imada10}.  While the phase diagrams of Fe-based systems have some commonalities, there are examples of good superconductors in this class that do not share them.  For example, unlike the ``canonical" phase diagram where superconductivity evolves via chemical doping from a magnetic parent compound, neither FeSe nor LiFeAs have  long range magnetism at ambient pressure, and require no chemical doping to become superconductors\cite{Fernandes22}.

Thus an  apparent overall lack of universality in the properties of the Fe-based systems themselves hinders the search for the underlying, presumed universal physical mechanism of superconductivity. Recently, however, it was pointed out by Miao  et al.\cite{Ding2018} that a quite large number superconducting Fe-based materials, from very weak to very strong superconductors, have at least one thing in common: a quite large value, of roughly 3.5, of the ratio of the maximum $T\rightarrow 0$ gap on the Fermi surface, $\Delta_{max}$, to $T_c$. To show this, values of the ratio  $\Delta_{max}/T_c$ for 16 materials with widely varying $T_c$'s were compiled from reported angle-resolved photoemission (ARPES) data.  In BCS theory for conventional superconductors,  this number $\Delta/T_c$ is 1.76, and is indeed referred to as a ``universal ratio" that emerges from the theory.
Of course, it is well known that both strong coupling effects beyond BCS\cite{Carbotte1990} and gap anisotropy\cite{Einzel2003} can increase this ratio.  Lead (Pb), for example, has a ratio $\Delta/T_c$ of about 2.25, and underdoped to optimally doped cuprates have quite
large ratios as well\cite{Chen2019}.  But it is less the size of the ratio that is remarkable in the Fe-based materials, than the apparently universality in a class of superconductors where so much is understood to depend on details of the electronic structure.  

Lee et al.\cite{Lee18} approached the problem in terms of a model reflecting  Hund's metal physics, including interactions mediated by nearly localized high energy spin modes.  These authors did not attempt to show that universality of the ratio would emerge from models reflecting the degree of variability in electronic structure of the Fe-based materials.  Instead, they studied the so-called ``$\gamma$-model" of a {\it one-band} superconductor with singular power law interactions described by a susceptibility $\chi\sim \Omega^{-\gamma}$, and showed that $\Delta_{max}/T_c$ could be increased substantially above the BCS value, and agreed with the Miao et al. value when $\gamma=1.2$.  Since this behavior is close to the observed behavior $\gamma$ of Fe-based superconductors at high energies, they  suggested that the observed quite large value of $\Delta_{max}/T_c$,  was  due to the  physics of a Hund's metal. Within the framework employed, they showed that large gap values were possible, but no real notion of a $\Delta_{max}/T_c$ robust against system details  emerged from these calculations.  
 
 Here we take a quite different approach, arguing that  the large $\Delta_{max}/T_c$ ratio of $\sim 3.5$ may in fact be most naturally explained in the framework of simple multiband weak-coupling theory. By weak-coupling, we mean that the  interactions within and among bands are assumed much smaller than the relevant  Fermi energies and the energies of the pairing bosons.
 We show, within a simple Suhl-Matthias-Walker (SMW) ansatz of constant  pairing interactions in band space\cite{Suhl1959},  that values of  $\Delta_{max}/T_c$ cluster close to a universal value characteristic of the effective mass ratio between the two bands of the model, {\it provided} the pairing is dominantly interband in nature. This occurs independent of band parameters and other system details. The measured value $\approx 3.5$ of the universal ratio does not emerge naturally unless a very large band mass ratio is assumed, however. As we show, this is {\it not } characteristic of the Fe-based superconductors.  
 The result is nonetheless quite striking, and we suggest on this basis that the dominant interband nature of the interactions, assumed in many discussions earlier but rather difficult to prove in practice, is the true essential feature of Fe-based superconductivity.  In the discussion section, we propose for more realistic multiorbital models a mechanism that may be analogous to the mass anisotropy in the simple 2-band case considered here. 
 
\section{Model}
\subsection{Two-band superconductivity}
\begin{figure}[h]
    \centering
    \includegraphics[width=0.49\textwidth]{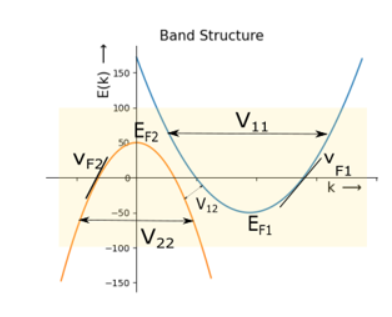}  
    \caption{The Band structure of our two band model for two different mass ratios. (a) mass ratio 1 and (b) mass ratio 2. $E_{f2} = 50 meV$ and $E_{f1}=50 meV$ are the fermi energies of the two bands. $V_{f2}$ and $V_{f1}$ are the fermi velocities of the two bands. $V_{11}$,$V_{22}$ and $V_{12}$ represent the interactions between the bands.The colored region is the energy cutoff $\omega_c = 100 meV$.       }
    \label{fig:Band structure}
\end{figure}

In contrast to Ref. \cite{Lee18}, we explore here under what circumstances a ``universal" ratio $\Delta_{max}/T_c$ might be expected in an ensemble of systems characterized by several bands, coupling constants, masses, etc.  
We begin with a simple two band model\cite{Suhl1959,Moskalenko59} with constant interactions within and between bands. To simulate Fe-based systems, we assume that one  is a hole band and the other an electron band, as shown in Fig. \ref{fig:Band structure}, with \begin{eqnarray}
\epsilon_1({\bf k}) &=& E_{F1}+{({\bf k}-{\bf k}_M)^2\over 2 m_1}\\
\epsilon_2({\bf k}) &=& E_{F2}-{k^2\over 2 m_2} 
\end{eqnarray}


We have modeled the intraband interactions $V_{11}$ and $V_{22}$, as well as interband interactions $V_{12}$, using the $\lambda$ interaction matrix-
\begin{gather*}
\begin{bmatrix}
\lambda_{11} & \lambda_{12} \\
\lambda_{21} & \lambda_{22} 
\end{bmatrix}
=
\begin{bmatrix}
m_1*V_{11} & m_2*V_{12}\\
m_1*V_{21} & m_2*V_{22}
\end{bmatrix},
\end{gather*}
where $m_1$ and $m_2$ are the band masses and positive interactions are taken to be attractive. 
The two-band BCS equations that determines $\Delta$ and $T_c$ are
\begin{eqnarray}
    \Delta_1=\lambda_{11}\Delta_{1}F(\Delta_1,T)+\lambda_{12}\Delta_{2}F(\Delta_2,T)\\
    \Delta_2=\lambda_{22}\Delta_{2}F(\Delta_2,T)+\lambda_{21}\Delta_{1}F(\Delta_1,T),
    \label{eq:BCS}
\end{eqnarray}
and the functions  $F_\alpha(\Delta,T)$ for the electron and hole bands $\alpha=1,2$ are 
\begin{eqnarray}
F_1\left[\Delta_{1}, T\right]&=& \int_{-E_{F1}}^{\omega_{D}} d \xi \frac{1}{2 \sqrt{\xi^{2}+\Delta_{1}^{2}}} \tanh \left(\frac{\sqrt{\xi^{2}+\Delta_{1}^{2}}}{2 T}\right)~~~\\
F_2\left[\Delta_{2}, T\right]&=&\int_{-\omega_D}^{E_{F2}} d \xi \frac{1}{2 \sqrt{\xi^{2}+\Delta_{2}^{2}}} \tanh \left(\frac{\sqrt{\xi^{2}+\Delta_{2}^{2}}}{2 T}\right),~~~
\end{eqnarray}
where we have taken the Fermi energies (band extrema) equal   $E_{F2}=|E_{F1}|\equiv \epsilon_F$ for simplicity\footnote{ We verified that results are  generally insensitive to changes in high-energy electronic structure. }, and measured energies relative to the chemical potential $\mu$,   $\xi =\epsilon_\alpha({\bf k}) - \mu$ as usual; hence all gaps open below $T_c$ around the fixed chemical potential $\mu$.\\ 

\begin{figure}[h!]
         \centering
         \includegraphics[width=0.49\linewidth]{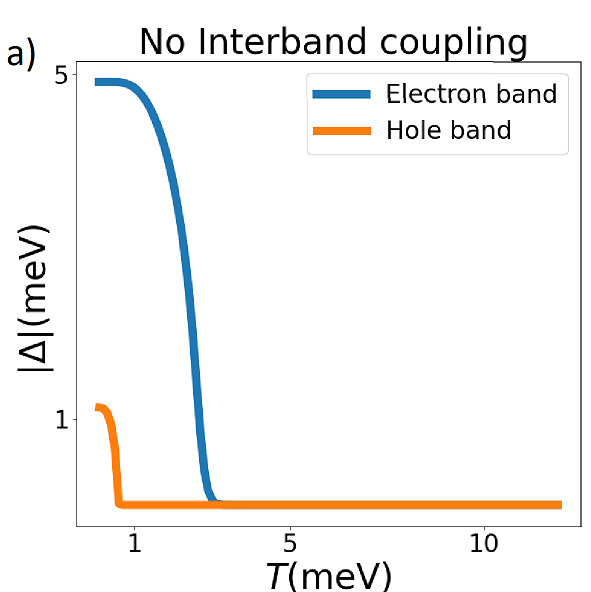}
         \includegraphics[width=0.49\linewidth]{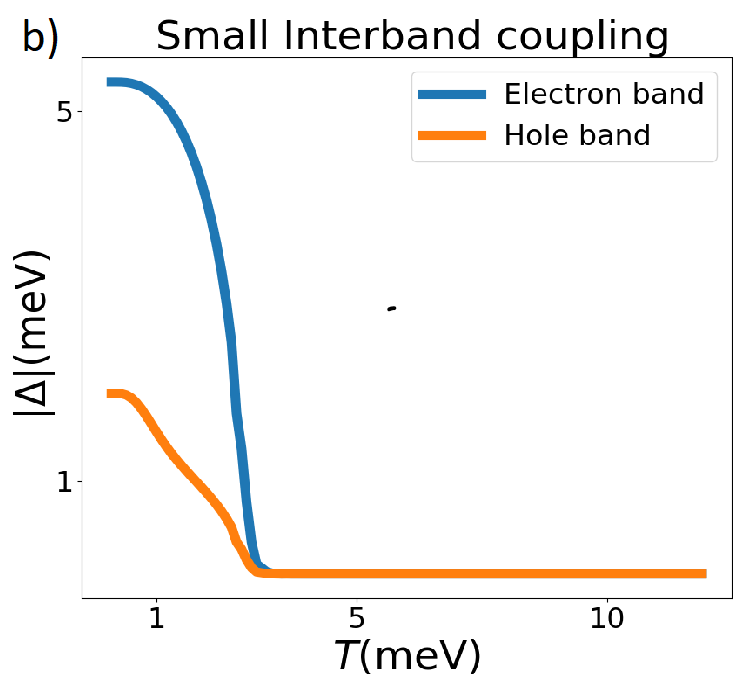}
         \includegraphics[width=0.45\linewidth]{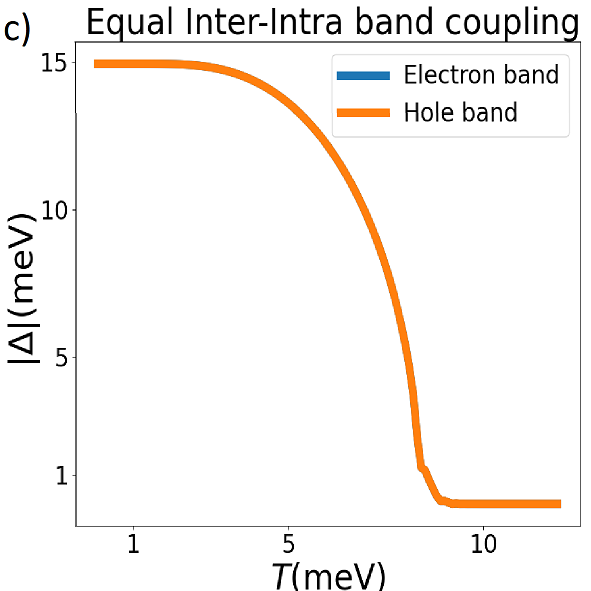}
         \includegraphics[width=0.49\linewidth]{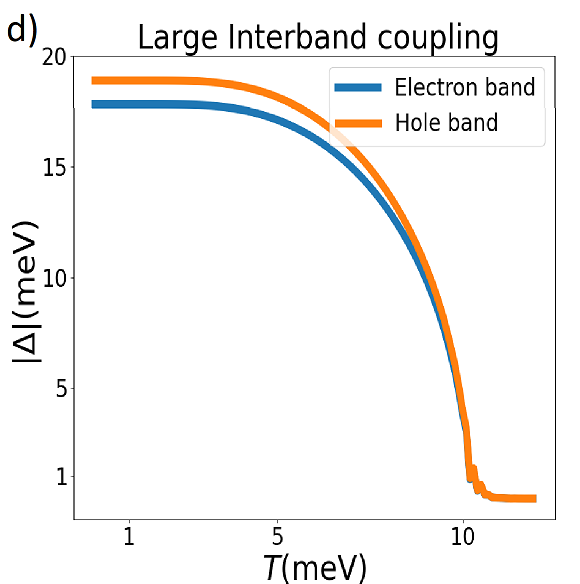}
         
         \caption{  $|\Delta|$ vs. $T$  in various regimes of intra vs intra-band pairing with  mass ratio 2.
         (a)  $\lambda_{11}=1.43$ $\lambda_{12/21}=0$; (b) $\lambda_{11}=1.43$ and $\lambda_{21}=0.143$; (c) $\lambda_{11} = \lambda_{21} = 1.43$; (d) $\lambda_{11}=1.14$ and $\lambda_{21}=1.43$ with  $\lambda_{22}= 2 \lambda_{11}$ and $\lambda_{12} = 2 \lambda_{21}$ everywhere.
        }
         \label{fig:examples}
 \end{figure}

We  performed a consistency check of our code for mass ratio 2 and different regimes of interband and intraband couplings. We observed from FIG. \ref{fig:examples} that when we have no interband coupling, the two $\Delta$ behave independently and give two different $T_c$. But as we progressively keep increasing the interband coupling strength $\Delta_1$ and $\Delta_2$ converge at the same $T_c$. Since mass of band 1 is larger then band 2 initially we have  $\Delta_1>\Delta_2$ but as the interband interaction increases we observe an increase in the $\Delta_2$ value until it finally becomes greater then $\Delta_1$ for interband larger than intraband interaction.
$\Delta_{max}$ is defined to be the maximum of $\Delta_{1}$,$\Delta_{2}$  at zero temperature. {  Note in Miao et al. \cite{Ding2018}, the maximum is also taken over the Fermi surface, but here we adopt an isotropic approximation for simplicity.  Although seldom discussed, it is clear that the $\Delta_{max}/T_c$ ratio can be quite large compared to 1.76, generally forcing the minimum gap ratio $\Delta_{min}/T_c$ at the same time to be considerably smaller. }

 To differentiate between the different regimes of inter vs intra band interactions we define a lograrithmic quantity-
\begin{equation*}
    \beta=\log{\lambda_{12}+\lambda_{21}\over \lambda_{11}+\lambda_{22}}
\end{equation*} 
 which parameterizes the degree of interband pairing. $\beta\rightarrow \infty$ is pure interband pairing, while $\beta\rightarrow -\infty$ is pure intraband pairing.



\subsection{``High-throughput" approach}

 \begin{figure*}[tb]
         \centering
 \includegraphics[width=0.49\linewidth,height=0.4\linewidth]{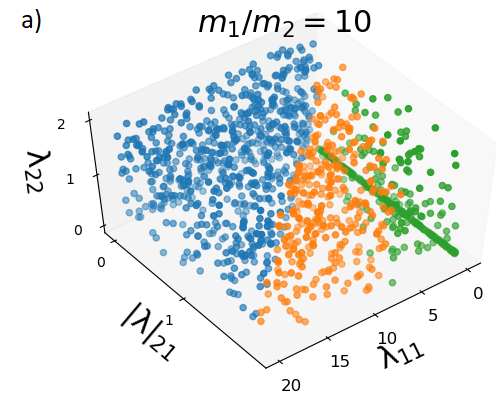}
\includegraphics[width=0.49\linewidth,height=0.4\linewidth]{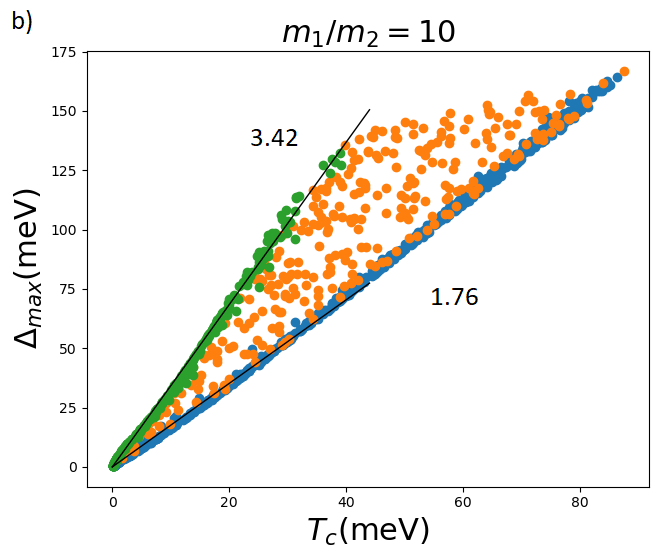}     
    \includegraphics[width=0.49\linewidth,height=0.4\linewidth]{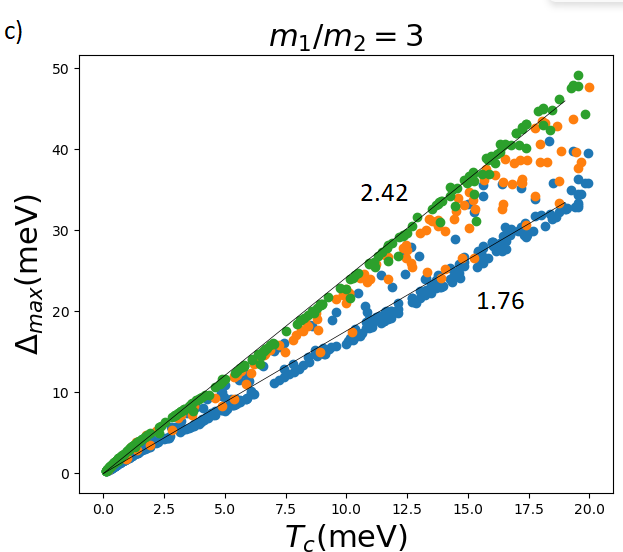}
   \includegraphics[width=0.49\linewidth,height=0.4\linewidth]{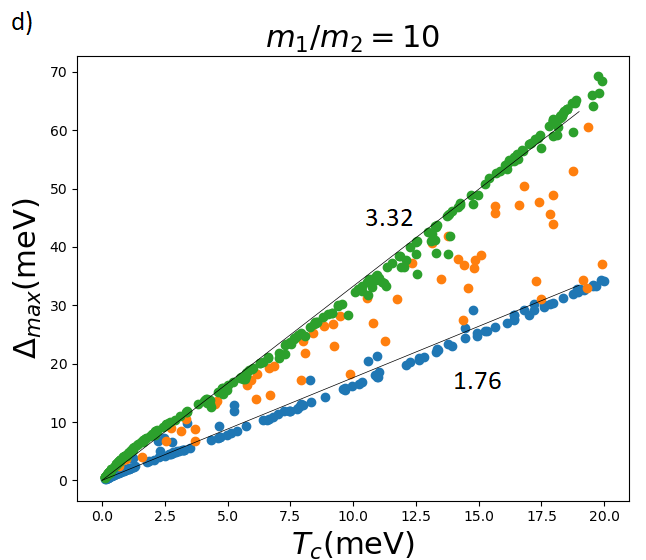}
         \caption{(a)  Uniform distribution of $\lambda_{ij}$ values used in ``high-throughput" search
         for fixed band mass ratio $m_1/m_2$=10.
         Blue point represents  $\beta <0$,  orange points  $0<\beta<1$ and  green points are for $\beta>1$, representing  three regimes of  interband and intraband potential ratios. (b) Calculated ratio of maximum gap to critical temperature,  $\Delta_{max}/T_c$ from 2-band BCS equations (\ref{eq:BCS}), using parameter sets represented in (a).  Note the clustering of points around weak coupling value 1.76 and strong coupling value 3.42.  (c),() Comparison of high throughput results for two different mass ratios(3 and 10)in the weak coupling regime. The upper and lower limits of $\Delta_{max}/T_c$ for large and small  $\beta$  are indicated by  thin solid lines with slopes given. 
         } 
         \label{Distribution}
 \end{figure*}

In a a materials class with many different members with quite different electronic structure, it is an appropriate starting point to assume them to have random values of the parameters determining $\Delta_{max}$ and $T_c$ within reasonable physical ranges.   First, we consider a uniform probablility distribution of the most important such parameters, namely the individual  interaction matrix elements $\lambda_{ij}$, as shown in Fig. \ref{Distribution}(a).
We have divided our study into three regimes based on the values of the elements of interaction matrix to help identify the qualitative underlying physics.  These include three ranges of $\beta$ identified in the Figure by their color coding,
strong interband coupling $\beta >1$ (green) ,
weak interband coupling $\beta<0 $ (blue) and
comparable interband and intraband coupling $0<\beta<1$ (orange). 

Fig.\ref{Distribution}(a) shows that our distribution is indeed uniform, i.e. not biased towards any particular regime of potentials.  { The way the points are distributed is like this. There are roughly equal number of points for $\beta<0$ and $\beta>0$ now since the regime for $\beta>0$ is divided in $\beta>1$ and $1>\beta>0$ the points are further divided between the two regimes.}
Fig.\ref{Distribution}(b) shows that such an evaluation give rise to $T_c$ values which are not in the physical range therefore when we look at the results of the high throughput approach for different mass ratios we will focus mainly at the physical values.

Our ``high throughput" approach then involves solving (\ref{eq:BCS}) for $T_c$ and the superconducting gaps for the various different values of $\lambda_{ij}$. We repeat the exercise for differing values of the Fermi energies, effective masses, and BCS pairing cutoffs to check what influences the $\Delta_{max}/T_c$ ratio.  
The overall coupling strength is adjusted to keep $T_c$ in a reasonable physical range for Fe-based superconductors.




\begin{figure}[h!]
     \centering
     
         \includegraphics[width=0.4\textwidth]{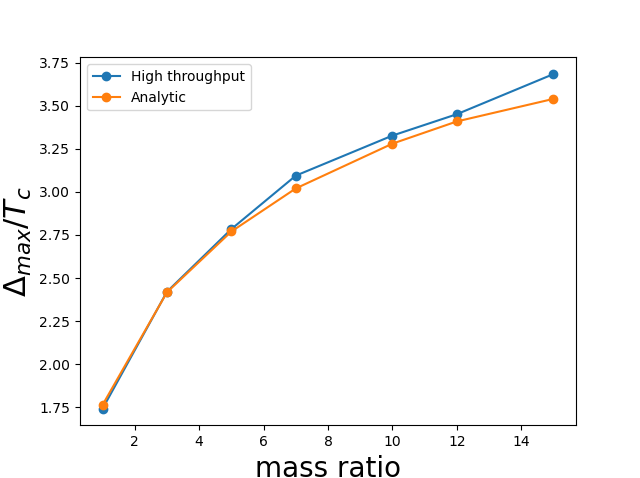}
         \caption{We compare the $\Delta_{max}/T_c$ ratios using our high throughput approach averaged over the $\beta>1$ regime and the exact solution for zero intraband coupling $V_{11}=V_{12}=0$. }
         \label{fig:analytic}
\end{figure} 
\section{Results}

In Fig.\ref{Distribution}, we show the results of our high throughput analysis. We consider two different mass ratios, $m_1/m_2 = 3, 10$ and report the value of $\Delta_{max}/T_c$ for different $\lambda_{ij}$.  In panel (b) we show the $\Delta_{max}/T_c$ values obtained for the full range of $\lambda$'s shown in panel (a) for the particular choice $m_1/m_2=10$.   Points are distributed between expected asymptotic ratio values: 1.76 for the weak negligible interband coupling case, as discussed above, and a value close to $\sqrt{m_1/m_2}=3.16$ for dominant interband coupling.   (see Fig. \ref{fig:analytic} below and accompanying discussion).    All points in between these asymptotes require numerical evaluation.    In (c)-(d), the distribution of $\Delta_{max}/T_c$ is shown for two different values of the mass ratios $m_1/m_2$, 3 and 10.  
Independent 
of the value of the mass ratio , we find some common features, namely clustering around the asymptotic values for small $\beta$ or large $\beta$. 
When the interband and intraband coupling are comparable, $1>\beta>0$, the values of $\Delta_{max}/T_c$ do not cluster and are instead scattered rather uniformly between the two asymptotes.
As seen in Fig. \ref{fig:examples}, in this regime were the inter/intra-band interactions are comparable, the two gaps $\Delta_1$ and $\Delta_2$ are close in magnitude and we observe a crossover from $\Delta_1/\Delta_2>1$ to $\Delta_1/\Delta_2<1$ when the interband becomes larger than the intraband interaction (for mass ratio $m_1/m_2>1$).
\\

In order to achieve a large ``universal" value of the  $\Delta_{max}/T_c$ ratio {\it within the current simplified framework}, it is clear that we must assume that all materials are dominated by interband pairing interactions,  $\beta >1$.  In this case,  we have a cluster of points near the upper limit for each mass ratio, so the $\Delta_{max}/T_c$ ratio is constant and larger than the BCS value for all ``samples".   The interesting observation is that the upper limit  increases with increasing  in the mass ratio, as shown in Figure \ref{fig:analytic},{  where we compare with the  interband-only solution which was obtained by setting the diagonal terms in the interaction matrix to zero.  This $T=0$ result is quite similar to the well-known exact result $|\Delta_1/\Delta_2|= \sqrt{m_2/m_1}$ that holds in this limit near $T_c$\cite{HKM2011}.
However, it requires a  mass ratio near 10 to achieve a $\Delta_{max}/T_c$ ratio of near the observed value of  3.5\cite{Ding2018}.  Such large mass differentiations are not observed in Fe-based systems, however.  In Table \ref{table}, we exhibit Fermi velocities on electron and hole bands in common units so that comparison across materials is straightforward.  Maximum values of the mass ratio 2-3 are much more typical than the larger values required within this theory.  Furthermore, mass ratios differ substantially from material to material, implying a breakdown of any possible universality that might have been present had all interactions been strongly interband in nature and all systems had a common mass ratio.  
    
{ Other parameters in the model were found to have much less significant impact on the $\Delta_{max}/T_c$ ratio.  For example, we have considered how these results change when the Fermi energies $E_{F1}$ and $E_{F2}$  are varied.  We find that the values of $\Delta_{max}$ and $T_c$ change but the ratio of $\Delta_{max}/T_c$ does not. The same was true for changes in the energy cutoff.  As expected in the weak-coupling limit, changes in the cutoff changed the values of $\Delta_{max}$ and $T_c$, but the ratio remained approximately the same for a given mass ratio}


\section{Discussion}
We have considered the effect of varying the most common physical quantities thought to influence a superconductor's critical temperature and gap function\cite{Fernandes22}, in  hopes of understanding if there is a natural explanation for the apparent ``universality" of $\Delta_{max}/T_c$ throughout the Fe-based family of materials.   The model considered here is obviously a vastly oversimplified description of these systems.  To name a few clear limitations, we have neglected the dynamics, momentum anisotropy, and orbital dependence of the interaction matrix, self-energy and vertex correction effects beyond BCS, the correct orbital-dependent electronic structure, spin orbit coupling, etc. Among these various limitations, let us single out gap anisotropy for special attention, since it is well-known that many of the Fe-based superconductors, including those analyzed by Miao et al\cite{Ding2018}, are extremely anisotropic, with accidental or symmetry-enforced gap nodes, while others are fully gapped.  For some  fixed interaction strength, making the gap more anisotropic enhances $\Delta_{max}/T_c$\cite{Einzel2003,Kogan2021}, suggesting naively that such variability should lead immediately to a breakdown in $\Delta_{max}/T_c$ universality.   On the other hand, such anisotropy is generally driven by variations in orbital content around the Fermi surface\cite{Maier09}, not included in the present approach.  
So it is not completely implausible that some more general measure of superconducting ``fitness"\cite{Ramires2016,Ramires2018} including orbital degrees of freedom might preserve the quantity $\Delta_{max}/T_c$ despite the apparent large variability of the anisotropy of the interaction across the Fe-based materials.

It is therefore tempting to speculate that it is the failure to include orbitals and their differentiated electronic correlations in the Fe-based systems which is the principal deficiency of the current model.   As has been pointed out in many places,  Hund's metal correlations drive a strong asymmetry between $d_{xy}$ and $d_{xz}/d_{yz}$ correlations, e.g. effective masses, a trend observed in nearly all the Fe-based materials\cite{Yi_review}.   It seems plausible that the  ``universal" correlation physics of the $d_{xy}-d_{xz/yz}$ splitting may play the role of the mass ratio in this simple model.  Investigations of this intriguing hypothesis are in progress.  

As a final sanity check, it is useful to compare the current analysis  to the more familiar case of MgB$_2$, to which the 2-band model has often been applied\cite{Kim2019}. This canonical diboride superconductor with $T_c=39K$ is generally considered to be an intraband-dominated multiband system with two main gaps driven by the electron-phonon interaction\cite{Budko2015}.  The average large gap in the system is empirically about 7 meV.  Experiments do not to our knowledge give reliable evidence for the gap {\it anisotropy}, but from EPW calculations\cite{Margine2013} the spread in the large gap value over the Fermi surface is $\pm 0.5$meV\footnote{The overall pairing scale in these calculations\cite{Margine2013} is well-known to be of order 20\% too high, but the relative measure of the gap anisotropy should be quite accurate.  } giving $\Delta_{max}/T_c$ of about 2.2, with (largest) effective mass ratios from de Haas - van Alphen measurements of about 3.5\cite{Elgazzar2002}.   This is roughly consistent with small $\beta$ values of $\Delta_{max}/T_c$ obtained from our simple model, see e.g. Fig. \ref{Distribution}(c).  Clearly the claimed Fe-based superconductor universal value of $\Delta_{max}/T_c$ of 3.5 represents a new kind of multiband superconducting physics compared to MgB$_2$, and we have suggested that -- up to a point -- our model indeed reproduces a universal ratio provided all systems are indeed in the interband dominated $\beta\gg 1$ limit.  The missing link within this framework to obtain actual agreement with the measured {\it value} $\Delta_{max}/T_c \simeq 3.5$ ratio is a large ({\cal O}(10)) effective mass ratio between bands, which is not empirically found in the Fe-based systems (Table \ref{table}).

\begin{table*}[t]
\centering
\begin{tabular}{ |p{3cm}|p{3cm}|p{3cm}|p{3cm}|p{3cm}| }
 \hline

 \hline
 Material & Hole-band  &Electron-band  &Average ratio&max ratio\\
 \hline
 FeTe$_{0.55}$ Se$_{0.45}$ \cite{PhysRevB.85.094506} & 95.54, 143.312    &N/A&  N/A &N/A\\
 Fe$_{1+y}$Se$_{x}$Te$_{1-x}$\cite{doi:10.1126/sciadv.1602372}&  166.66,166.66,125  & N/A   &N/A&N/A\\
 LiFeAs\cite{PhysRevB.94.201109}  &31.85,127.38,0 & 127.38,95.54& 1.4&4\\
 Ba$_{0.1}$K$_{0.9}$Fe$_{2}$As$_{2}$\cite{PhysRevB.88.220508} &84.92,99.04,113.23 & N/A& N/A&N/A\\
 KFe$_{2}$As$_{2}$\cite{article}& 60.51,67.2  & 37.26,34.6&1.7&1.94\\
 NaFe$_{0.95}$Co$_{0.05}$\cite{PhysRevB.84.064519}& 24.06  & 34.4,80.19   &2.38&3.33\\
 BaFe$_{2}$As$_{2}$\cite{PhysRevLett.105.087001}&56.78,113.56  & 115&1.35&2.02\\
 BaFeRu$_{0.35}$As\cite{PhysRevLett.105.087001}&248.4,248.4&344.9,172.45&1.04&1.39\\

 \hline
\end{tabular}
\caption{Fermi velocities (meV$-a$) of electron and hole bands evaluated from ARPES data of different Iron based Superconductors. The average mass ratio and the maximum mass ratio were evaluated by the ratio of electron and hole Fermi velocity.} 
\label{table}
\end{table*}
~
\vskip .2cm ~
\section{Conclusions}  To investigate the possible existence of ``universality" of the ratio $\Delta_{max}/T_c$ in iron-based superconductors as claimed by Miao et al.\cite{Ding2018}, we  solved the weak-coupling 2-band equations of superconductivity in a ``high-throughput" approach where interaction parameters $\Lambda_{ij}$, Fermi energies, and masses of electron and hole bands were varied over large but physical ranges of parameters, and the maximum gap and $T_c$ were determined.
We found that for a given mass ratio there is a distribution of $\Delta_{max}/T_c$ values between the 'BCS' value(1.76) and an upper limit which depends on the mass ratio between electron and hole bands. Although the various parameters are distributed uniformly, there is substantial clustering about the two lines which correspond to the intraband and interband dominated limits.   In the intermediate regime we have a scatter for many different values. The upper (interband interaction-dominated) limit increases with increasing  mass ratio according to the expected square root behavior.  We find that the 2-band weak-coupling model does {\it not} display a universal
$\Delta_{max}/T_c$ value as observed in ARPES experiments on Fe-based superconductors, but based on the systematics of the simple 2-band case we proposed a study including electronic correlations that may represent a promising way forward.
\section{Acknowledgements}

P.J.H. is grateful to G. Kotliar for bringing Ref. \cite{Ding2018} to his attention, and the authors thank L. Fanfarillo, V. Kogan, and R. Prozorov  for useful discussions.
 S.P. and P.J.H. were supported by the U.S. Department of Energy under Grant No. DE-FG02-05ER46236.

\bibliography{references}

\end{document}